\documentclass{PoS}

\newcommand{\mpic}{M_\pi^{crit}}

\title{Universality in QCD and Halo Nuclei }

\ShortTitle{Universality in QCD and Halo Nuclei }

\author{\speaker{H.-W. Hammer}%
        \thanks{This work was supported by the BMBF under contract No.
         06BN411.}\\
        Helmholtz-Institut f\"ur Strahlen- und Kernphysik (Theorie) and
        Bethe Center for Theoretical Physics, University of Bonn, Germany\\
        E-mail: \email{hammer@itkp.uni-bonn.de}}

\abstract{
Effective Field Theory (EFT) provides a powerful framework to exploit 
a separation of scales in order to perform systematically improvable, 
model-independent calculations. We apply this method to strongly
interacting quantum systems with short-range interactions and
large scattering length. Such systems display remarkable universal 
properties which include a geometric spectrum of shallow three-body 
states called "Efimov states" and log-periodic dependence of scattering 
observables on the scattering length.
We review the EFT for large scattering length and some 
of its applications in the physics of cold atoms and nuclear physics. 
In particular, we discuss the possibility of an infrared limit
cycle in QCD and the extension of the EFT to halo nuclei.
}

\FullConference{8th Conference Quark Confinement and the Hadron Spectrum\\
		 September 1-6, 2008\\
		 Mainz. Germany}

\begin{document}

{\it Introduction.}
The Effective Field Theory (EFT) approach provides a powerful framework 
that exploits the separation of scales in physical systems. 
Only low-energy (or long-range) degrees of freedom are included explicitly, 
while everything else is parametrized in terms of the most general
local contact interactions. 
Thus, the EFT describes universal low-energy physics independent of
detailed assumptions about the short-distance dynamics. 
Physical observables can be described in a controlled expansion in 
powers of $kl$, where $k$ is the typical momentum and
$l\sim r_e$ is the characteristic low-energy length scale of the system. 
We focus on applications of EFT to few-body systems
with large S-wave scattering length $|a| \gg l$. 
For a generic system, the scattering length
is of the same order of magnitude as the low-energy length scale $l$.
Only a very specific choice of the parameters in the underlying theory 
(a so-called {\it fine tuning}) will generate a large scattering 
length $a$. The fine tuning can be accidental or it can be 
controlled by an external parameter. 
Examples with an accidental fine tuning are the S-wave scattering 
of nucleons or of $^4$He atoms. The scattering length of alkali atoms close 
to a Feshbach resonance can be tuned experimentally by adjusting the
external magnetic field.
At very low energies these systems behave similarly and
show universal properties associated with large $a$ \cite{BrH04}.
We start with a brief review of the EFT for 
few-body systems with large $a$ and then discuss some 
applications in nuclear and atomic physics. 

\begin{figure}[t]
\bigskip
\centerline{\includegraphics*[width=12cm,angle=0]{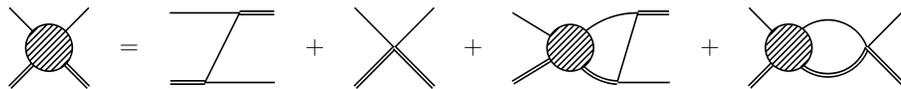}}
\caption{Integral equation for the particle-dimer scattering amplitude.
A single (double) line indicates a particle (full dimer) propagator.}
\label{fig:ineq}
\end{figure}

{\it Three-body system with large scattering length.}
We consider a two-body system of bosonic particles
with large S-wave scattering length $a$ and mass $m$. 
The generalization to fermions is straightforward.
For momenta $k$ of the order of
the inverse scattering length $1/|a|$, the problem is nonperturbative 
in $ka$. The exact two-particle scattering amplitude can be obtained
analytically by summing the so-called {\it bubble diagrams} with 
a 2-body contact interaction.
The resulting amplitude reproduces the leading order of the 
effective range expansion for the particle-particle
scattering amplitude:~$f_{AA}(k)=(-1/a -ik)^{-1}\,,$
where the total energy is $E=k^2/m$. 
If $a>0$, $f_{AA}$ has a pole at $k=i/a$ corresponding
to a shallow dimer with binding energy $B_2=1/(ma^2)$. 
Higher-order derivative interactions are perturbative and generate
corrections suppressed by powers of $\ell/|a|$.
We now turn to the 3-body system.
At leading order, the particle-dimer scattering amplitude is given by the 
integral equation shown in  Fig.~\ref{fig:ineq}. 
The inhomogeneous term consists of the one-particle exchange
and the 3-body contact interaction. The integral equation simply sums
these diagrams to all orders.
An ultraviolet cutoff $\Lambda$ must be introduced
in order to regulate the loop integrals 
involved. This cutoff guarantees that the integral equation
has a unique solution. All physical observables, however, must be 
independent of $\Lambda$, which determines the behavior of 
the 3-body contact interaction $H$ as a 
function of $\Lambda$ \cite{BHvK99}:
\begin{eqnarray}
H (\Lambda) = {\cos [s_0 \ln (\Lambda/ \Lambda_*) + \arctan s_0]
\over \cos [s_0 \ln (\Lambda/ \Lambda_*) - \arctan s_0]}\,,
\label{H-Lambda}
\end{eqnarray}
where $s_0=1.00624$ is a transcendental number and $\Lambda_*$
is a 3-body parameter introduced by dimensional transmutation. 
It cannot be predicted by the EFT and must be 
determined from a 3-body observable.
Note also that $H (\Lambda)$ is periodic and runs on
a limit cycle. When $\Lambda$ is increased by a factor of
$\exp(\pi/s_0)\approx 22.7$, $H (\Lambda)$ returns to its original 
value. In summary, two parameters are required to specify a 
3-body system at leading order in $l/|a|$:
they may be chosen as the scattering length $a$ (or equivalently
$B_2$ if $a>0$) and the 3-body parameter $\Lambda_*$ \cite{BHvK99}. 
This universal EFT confirms and extends the universal predictions for
the 3-body system derived by Efimov including
the {\it Efimov effect}, the accumulation of infinitely many 3-body 
bound states at threshold as $a\to\pm\infty$ \cite{Efi71}.

\begin{figure}[t]
\centerline{\includegraphics*[width=6.3cm,angle=0]{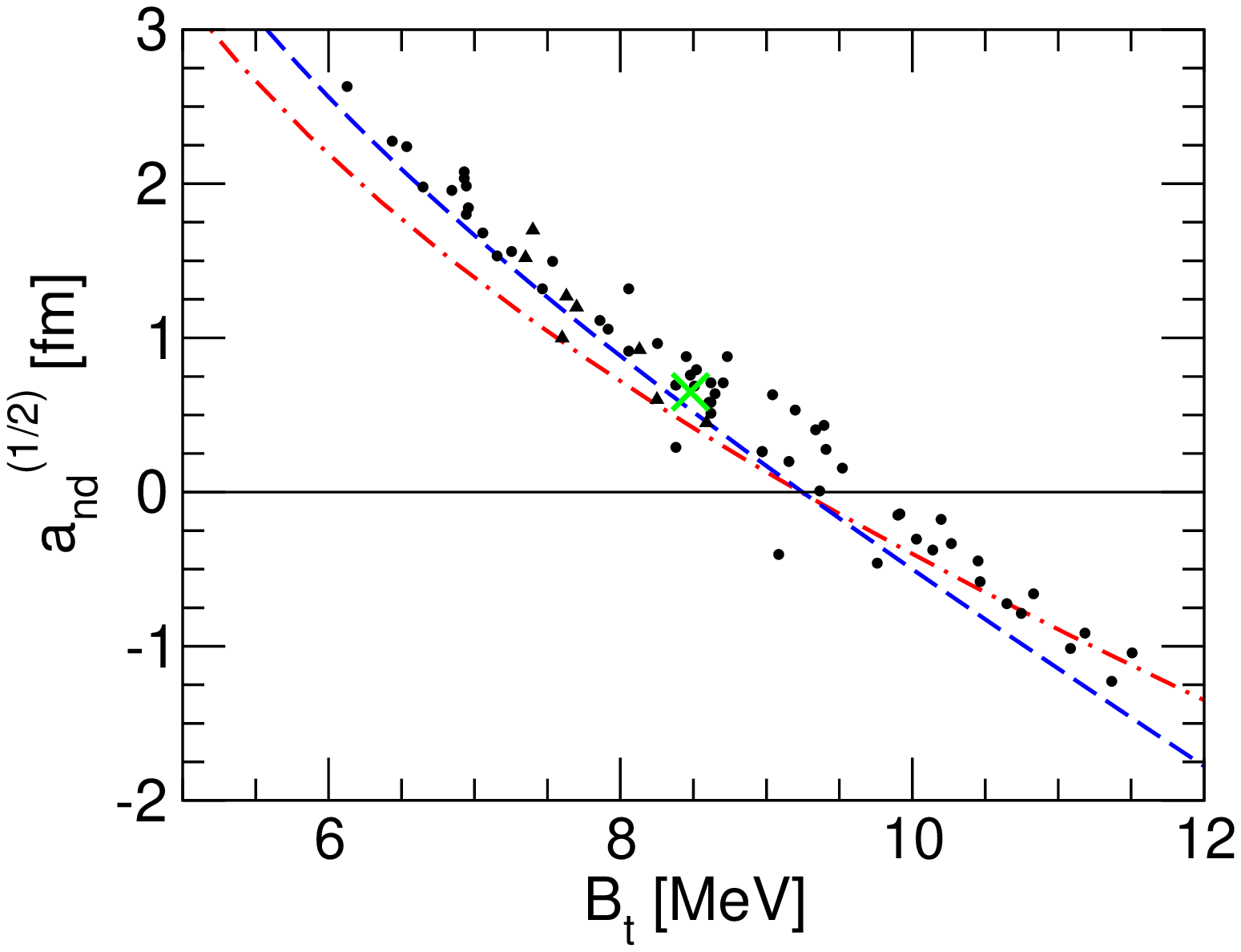}
\qquad\includegraphics*[width=6.7cm,angle=0]{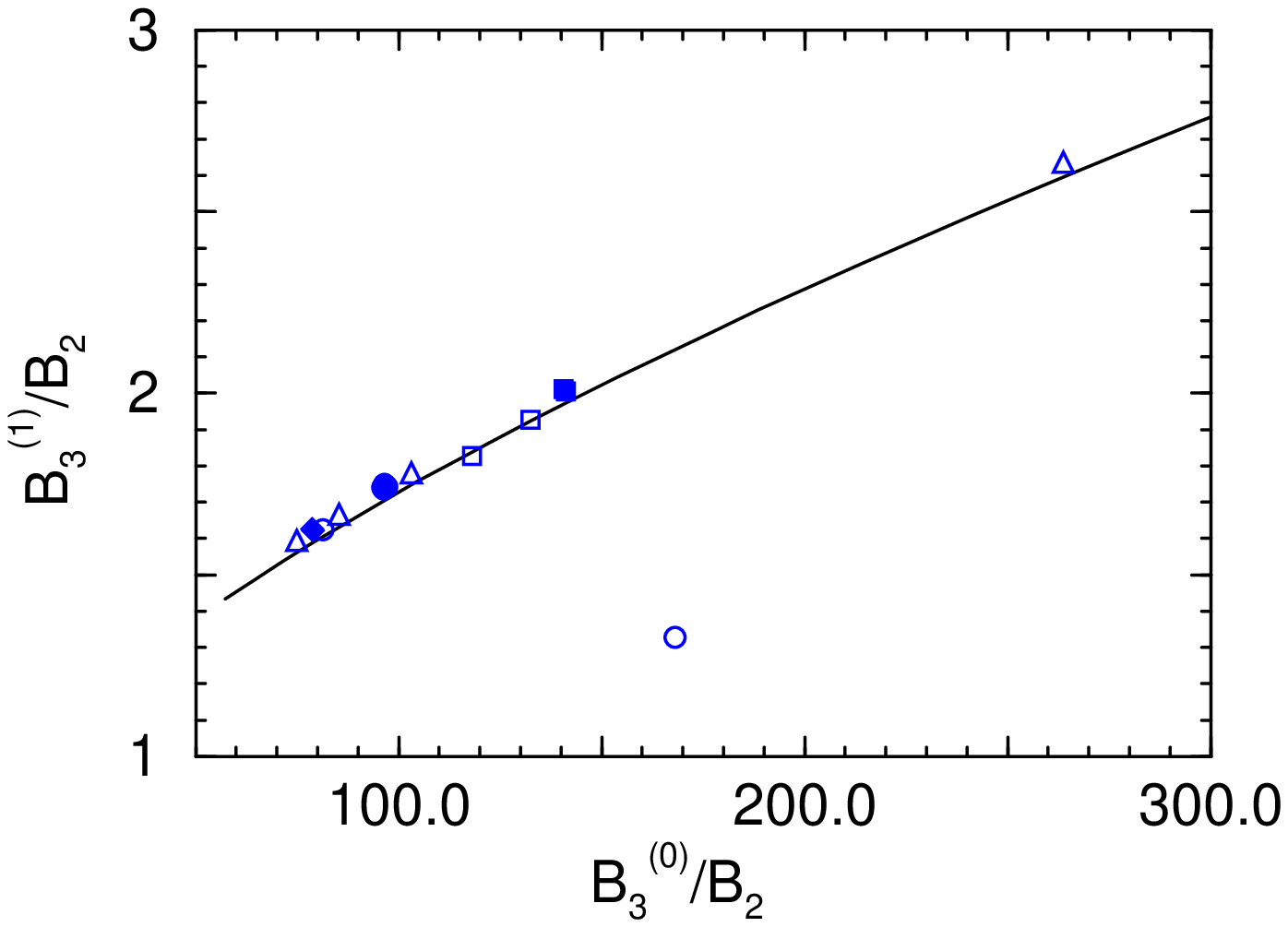}}
\caption{Universal correlations between the triton binding energy and 
the spin-doublet neutron-deuteron scattering length (left panel) and
between $^4$He trimer ground and excited state energies $B_3^{(0)}$ and 
$B_3^{(1)}$ (right panel).}
\label{fig:scale3}
\end{figure}

{\it Universal correlations.}
Since up to corrections of order $l/|a|$,
low-energy 3-body observables depend on $a$ and $\Lambda_*$ only, 
they obey non-trivial scaling relations. If dimensionless combinations
of such observables are plotted against each other, they must fall close
to a line parametrized by $\Lambda_*$ \cite{BrH04,BHvK99,Efi71}.
These correlations are purely driven by the large scattering length
and are independent of the mechanism responsible for it.
In Fig.~\ref{fig:scale3}, we show two examples of such universal
correlations \cite{BrH04}.
In the left panel, we show the Phillips line, a correlation
between the triton binding energy and the doublet neutron-deuteron
scattering length. In the right panel, we show the correlation between 
the $^4$He trimer ground and excited state energies $B_3^{(0)}$ and 
$B_3^{(1)}$.
The data points show calculations using various
interaction potentials. Since these potentials have approximately the same
scattering length but include different short-distance physics,
the points on this line correspond to different values of $\Lambda_*$.
The small deviations of the potential model calculations are mainly due to
effective range effects. They are suppressed by $r_e/|a|$
and can be calculated at next-to-leading order.
The extension of this EFT to the 4-body system requires no new
4-body parameter at leading order in $l/|a|$ \cite{Platter:2004qn}. 
Consequently, the universal correlations persist in the 4-body
system.

\begin{figure}[tb]
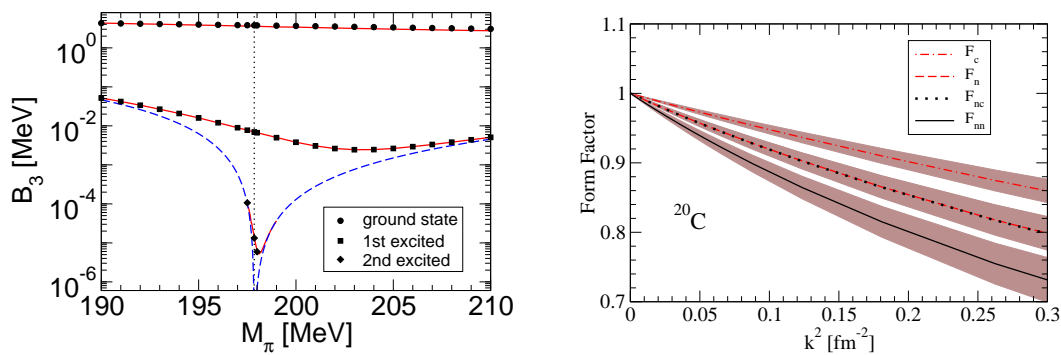

\centerline{\includegraphics*[width=6.7cm,angle=0,clip=true]{B3_spectrum3.eps}
\qquad\includegraphics*[width=6.4cm,angle=0]{figFF20C3.eps}}
\caption{Left panel: Binding energies of the triton ground and first two
excited states in the critical region as a function of $M_\pi$.
Right panel: The one- and two-body matter density form factors 
${\mathrm F}_{c}$, ${\mathrm F}_{n}$, ${\mathrm F}_{nc}$, and 
${\mathrm F}_{nn}$ with 
leading order error bands for the ground state of $^{20}$C as a function
of the momentum transfer $k^2$.}
\label{fig:spec.halo}
\end{figure}

{\it An infrared renormalization group limit cycle in QCD.}
Nuclear phenomena can be described within a chiral EFT which 
has the explicit dependence on the quark masses \cite{Epelbaum:2008ga}.
It has been used to study the quark mass dependence of nuclear
forces \cite{Beane:2002xf,Epelbaum:2002gb}. 
The extrapolation of the S-wave nucleon-nucleon scattering lengths
$a_t$ (spin triplet) and $a_t$ (spin singlet)
to larger values of $M_\pi$ predicts that $a_t$ diverges and 
the deuteron becomes unbound at a critical 
value in the range 170 MeV $< \, M_\pi \, <$ 210 MeV.
It is also predicted that $a_s$ is likely to diverge
and the spin-singlet deuteron becomes bound 
at some critical value of $M_\pi$ not much larger than 150 MeV. 
Based on this behavior it was conjectured that 
one should be able to reach the critical point by varying  
the up- and down-quark masses $m_u$ and $m_d$ independently
because the spin-triplet and spin-singlet channels have different isospin
\cite{Braaten:2003eu}.
In this case, the triton would display the Efimov effect which
corresponds to the occurence of an infrared limit cycle in QCD.
In Refs.~\cite{Epelbaum:2006jc,Hammer:2007kq}, 
the properties of the triton around
the critical pion mass were studied for one particular solution with 
a critical pion mass $\mpic=197.8577$ MeV. 
The binding energies of the triton and the first
two excited states in the vicinity of the limit cycle
were calculated for this scenario in chiral EFT.
They are given in the left panel of Fig.~\ref{fig:spec.halo}
by the circles, squares, and diamonds. The dashed lines indicate the
neutron-deuteron ($M_\pi \leq \mpic$) and 
neutron-spin-singlet-deuteron ($M_\pi \geq \mpic$) thresholds
where the 3-body states become unstable. Directly at 
the critical mass, these thresholds coincide with the 3-body
threshold and the triton has infinitely many excited states.
The solid lines are leading order calculations in the universal EFT using 
the pion mass dependence of the nucleon-nucleon scattering lengths
and one triton state from chiral EFT as input.
Both calculations agree very well and the universal EFT has also been 
used to calculate scattering \cite{Hammer:2007kq}.
The binding energy of the triton ground state varies only weakly over 
the whole range of pion masses. The excited states, however, are
more influenced by the thresholds and vary strongly.
Whether the limit cycle can be realized in QCD 
can only be answered definitely by directly solving QCD. In particular,
one would like to know whether this can be achieved
by approriately tuning the quark masses in a Lattice QCD simulation
\cite{Wilson:2004de}.

{\it Halo Nuclei.}
Halo nuclei can be described by extensions of the universal EFT. 
One can assume the core to be structureless and treats 
the nucleus as a few-body system of the core and the valence nucleons.
Corrections from the structure of the core appear in higher orders and
can be included in perturbation theory. 
A new facet is the appearance of resonances as in the neutron-alpha system
\cite{Bertulani:2002sz}.
The first application of effective field theory methods to halo nuclei 
was carried out in Refs.~\cite{Bertulani:2002sz,Bedaque:2003wa}, where the 
$n\alpha$ system (``$^5$He'') was considered. It was found that 
for resonant P-wave interactions both the scattering length and effective 
range have to be resummed at leading order. At threshold, however, only 
one combination of coupling constants is fine-tuned and the EFT becomes
perturbative. More recent studies have 
focused on the consistent inclusion of the Coulomb interation in two-body 
halo nuclei such as the $p\alpha$ and $\alpha\alpha$ 
systems \cite{Higa:2008rx,Higa:2008dn}. 

Three-body halo nuclei composed of a core and two valence neutrons
have the possibility to exhibit excited states due to
the Efimov effect \cite{Federov:1994cf}.
A comprehensive study of S-wave halo nuclei in EFT including
structure calculations with error estimates was recently carried out 
in Ref.~\cite{Canham:2008jd}.
Currently, the only possible candidate for an excited Efimov state is $^{20}$C,
which consists of a core nucleus with spin and parity quantum numbers
$J^P=0^+$ and two valence neutrons.  The value 
of the $^{19}$C energy, however, is not known well enough to 
make a definite statement about the appearance of an excited state
in $^{20}$C. The structure of halo nuclei can also be calculated
in the halo EFT. As an example, we show the 
various one- and two-body matter density form factors 
${\mathrm F}_{c}$, ${\mathrm F}_{n}$, ${\mathrm F}_{nc}$, and 
${\mathrm F}_{nn}$ (for a definition, see \cite{Canham:2008jd})
with leading order error bands for the ground state of $^{20}$C as a function
the momentum transfer $k^2$ in the right panel of Fig.~\ref{fig:spec.halo}.
The theory breaks down for momentum transfers of the order of the pion-mass 
squared ($k^2\approx 0.5$ fm$^{-2}$). 
{}From the slope of the form factors one can extract the radii:
${\mathrm F}(k^2) =  1 
- {1 \over 6} k^2 \left\langle r^2 \right\rangle + \ldots$\ .
Experimental information on these radii is available for
some halo nuclei. 
For the neutron-neutron radius of the Borromean halo nucleus $^{14}$Be
for example, the leading order halo EFT result is
$\sqrt{\langle r_{nn}^2\rangle}=4.1 \pm 0.5$~fm.
The value $\sqrt{\langle r_{nn}^2\rangle_{exp}}
=5.4 \pm 1.0$~fm was obtained from 
3-body correlations in the dissociation of $^{14}$Be using
intensity interferometry and Dalitz plots 
\cite{Marques:2001pe}.  Within the errors there is good agreement
between both values. 
Results for further halo nuclei are given in Ref.~\cite{Canham:2008jd}.
With upcoming experiments much more knowledge can be obtained 
about the structure of these intriguing systems as well as discovering 
whether they show universal behavior and excited Efimov states.

\end{document}